\documentclass[conference,a4paper]{APSIPA2021}
\usepackage{multirow}
\usepackage{graphicx}
\usepackage{amsmath}
\usepackage[psamsfonts]{amssymb}
\usepackage{amsxtra}
\usepackage{threeparttable}
\usepackage{booktabs}
\usepackage{subfigure}

\begin{document}

\title{Pay Attention to Hard Trials}

\author{%
\authorblockN{%
Lantian Li\authorrefmark{2}\authorrefmark{1},
Di Wang\authorrefmark{3}\authorrefmark{1},
Dong Wang\authorrefmark{1}
}
\authorblockA{%
\authorrefmark{1}
Center for Speech and Language Technologies, BNRist, Tsinghua University, China \\ }
\authorblockA{%
\authorrefmark{2}
School of Artificial Intelligence, Beijing University of Posts and Telecommunications, China \\ }
\authorblockA{%
\authorrefmark{3}
Key Laboratory of China's Ethnic Languages and Information Technology of Ministry of Education, \\ Northwest Minzu University, China \\
E-mail: \{lilt,wangdi\}@cslt.org; wangdong99@mails.tsinghua.edu.cn}
}

\maketitle
\thispagestyle{empty}

\begin{abstract}
Performance of speaker recognition systems is evaluated on test trials.
Although as crucial as rulers for tailors, trials have not been carefully treated so far,
and most existing benchmarks compose trials by naive cross-pairing. In this paper, we argue that
the cross-pairing approach produces overwhelming easy trials, which in turn leads to
potential bias in system and technique comparison.
To solve the problem, we advocate more attention to hard trials.
We present an SVM-based approach to identifying hard trials and
use it to construct new evaluation sets for VoxCeleb1 and SITW.
With the new sets, we can re-evaluate the contribution of some recent technologies.
The code and the identified hard trials will be published online at \emph{http://project.cslt.org}.
\end{abstract}

\section{Introduction}

The goal of automatic speaker verification (ASV) is to testify the claimed identity of a speech segment~\cite{campbell1997speaker,reynolds2002overview,hansen2015speaker}.
Recently, ASV has gained great popularity in a wide range of applications including access control, forensic evidence provision, and user authentication in telephony banking.
Traditional ASV methods are based on statistical models, for instance GMM-UBM~\cite{Reynolds00}, JFA~\cite{Kenny07} and i-vector model~\cite{dehak2011front},
while recent research focuses on deep learning methods, represented by the d-vector model~\cite{ehsan14,li2017deep} and x-vector model~\cite{snyder2018xvector,okabe2018attentive}.

The technical advance has led to significant performance improvement on numerous benchmarks.
For example, with a bunch of state-of-the-art (SOTA) architectures/techniques, such as ResNet topology~\cite{he2016deep,desplanques2020ecapa,zhou2021resnext},
attentive pooling~\cite{zhu2018self,okabe2018attentive}, angular margin loss~\cite{wang2018additive,deng2019arcface} and
score normalization \& calibration~\cite{kenny2010bayesian,matejka2017analysis},
researchers have reported equal error rates (EER) less than 1.0\% on SITW~\cite{mclaren2016speakers} and VoxSRC test sets~\cite{chung2019voxsrc,nagrani2020voxsrc}.

The continuously reduced EER reflects the technical progress to some extent;
however, the EER values are dubious, as practitioners often observe drastic
performance degradation in real deployments.
One reason is that the test data of these benchmarks are lack of real-world complexity,
hence simpler than data in real scenarios.
Actually, in a more challenging CN-Celeb test set~\cite{fan2020cn,li2020cn},
the EER values are much higher, as demonstrated by the results~\cite{tan22odyssey,zhang22codyssey,kang22dodyssey} of the recent CNSRC challenge organized on Odyssey 2022~\cite{cnsrc2022plan,cnsrc2022tech}.

Another reason is that the test trials in the benchmark tests involve a large proportion of easy trials, especially easy negative trials.
This is because the test trials are usually designed by cross-pairing in most benchmarks.
Specifically, suppose there are $N$ speakers in the test set, each having $K$ utterances, cross-pairing selects any two utterances (or the corresponding embedding vectors) as a trial and
labels it as positive if the two utterances are from the same speaker or negative if they are from
different speakers.
This leads to $NK(K-1)$ positive trials and $N(N-1)K^2$ negative trials.
Our argument is that most of the negative trials are easy and thus do not pose any challenge to the recognition system.
An immediate consequence is that EER results obtained with such
trial sets are over-optimistic and incapable of representing the true performance of deployed systems,
as we have analyzed in~\cite{li2022cp}.
Another consequence, perhaps more subtle and more serious, is that with the overwhelming easy trials, system comparison could be biased. We will show
that involving unbalanced easy trials causes a drifted trade-off between false acceptance and false rejection.

To address the issue, we advocate focusing on hard trials, i.e., trials that have not been well judged by existing techniques.
We will answer two basic questions in the following section:
(1) Why do we need to identify hard trials, or what is the problem if a large amount of easy trials are involved in system comparison?
(2) How to define and identify hard trials?
Once the answers are clear, Section~\ref{sec:exp} will present the identification results
on VoxCeleb1 and SITW, two famous test sets in speaker recognition.

\section{Theory and method}

\subsection{Why isolate hard trials}

What is the impact on system comparison when vast easy trials are involved in the test?
Firstly of all, it is straightforward to notice the \emph{absolute} difference on evaluation metrics (e.g., EER)
will be diminished with a large amount of easy trials involved,
as the decision error rates always converge to zero if there are numerous easy trials.
However, the \emph{relative advantage/disadvantage} could remain when comparing two systems.
If that is true, then involving easy trials would not change the conclusion in system comparison and
thus the existing cross-pairing scheme for trials design would not cause too many problems.

We conduct the analysis by adding easy \emph{negative} trials. Specifically, suppose there are
$N$ negative trials, and $m$ trials are correctly identified by System I and $m + k$ trials are correctly identified by System II.
Now let's add $s$ easy negative trials to the test and suppose they are all correctly recognized by the two systems.
It is easy to compute the relative change on false acceptance rate (rFAR) as follows:

\[
rFAR = \frac{\frac{N - m}{N + s} - \frac{N - m - k}{N + s} }{\frac{N - m }{N + s }} = \frac{k}{N - m}.
\]

\noindent Note that the resultant $rFAR$ is independent of $s$, the number of extra easy trials.
This implies that no matter how many easy trials are added to the test set, the relative FAR change remains the same.
In other words, it seems that identifying hard trials is not necessary.

However, the above derivation assumes that the decision threshold does not change.
This is not the case in performance evaluation.
For instance, when computing EER, the threshold is automatically selected to balance false acceptance (FA) and false rejection (FR).
The situation is the same when computing minDCF and other metrics that take FA/FR trade-off into account.
To show how easy trials impact system comparison in terms of relative EER reduction,
we conduct a simulation study.

\begin{figure}[htb!]
  \centering
  \includegraphics[width=0.88\linewidth]{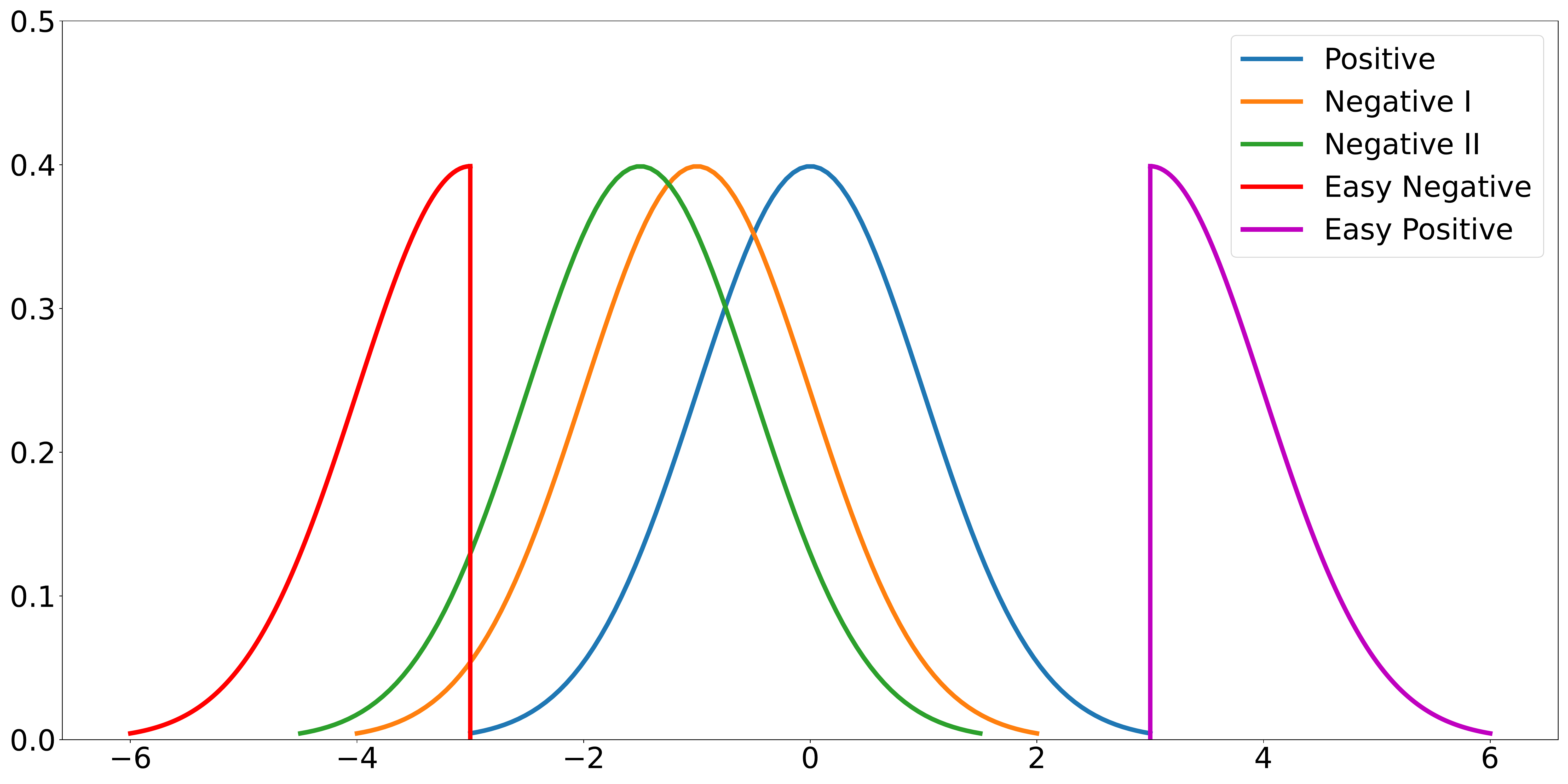}
  \caption{Data distributions used in the simulation study.}
  \label{fig:gauss2}
\end{figure}

Specifically, we randomly sample 10,000 samples from $N(0,1)$ as scores of positive trials,
and 10,000 samples from $N(-1,1)$ plus 10,000 samples from $N(-1.5, 1)$ as scores of negative trials
produced by system I and system II, respectively.
We then gradually add easy negative trials from $N(-3, 1)$ and/or scores of easy positive trials from $N(3,1)$.
In our experiment, we add 500 easy trials for each time and conduct the addition operation for 1000 times.
Fig.~\ref{fig:gauss2} presents an illustration for the distributions of the data samples.

\begin{figure}[htb!]
  \centering
    \subfigure[Negative]{\includegraphics[width=0.48\linewidth]{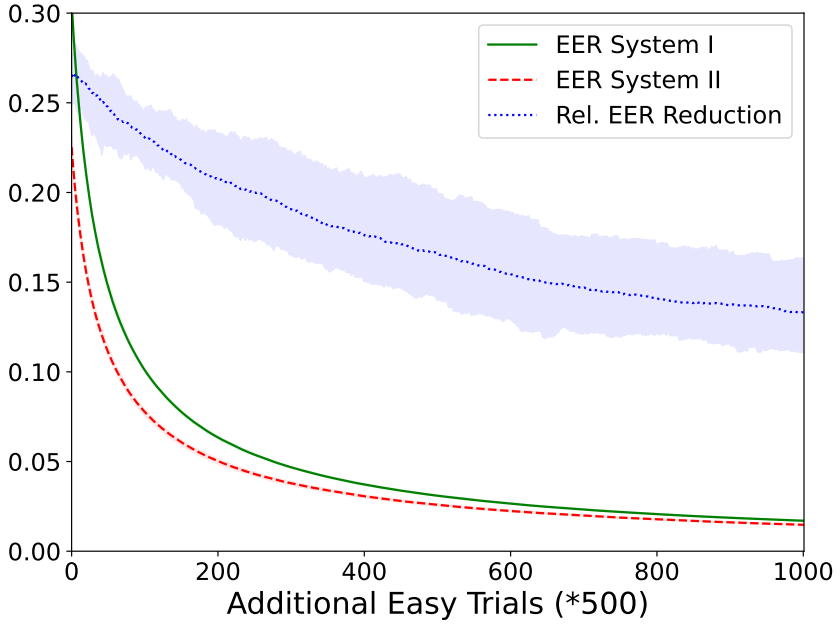}}
    \subfigure[Positive]{\includegraphics[width=0.48\linewidth]{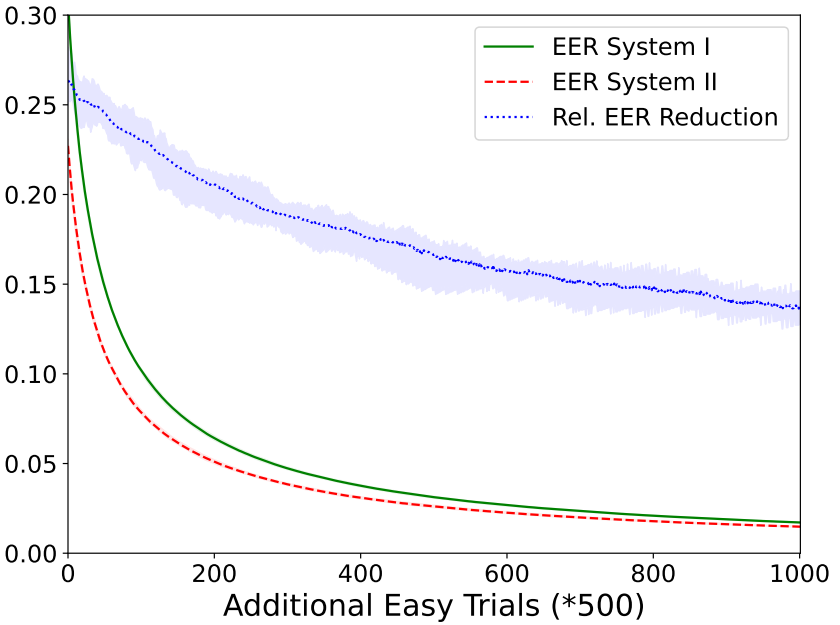}}
    \subfigure[Both]{\includegraphics[width=0.48\linewidth]{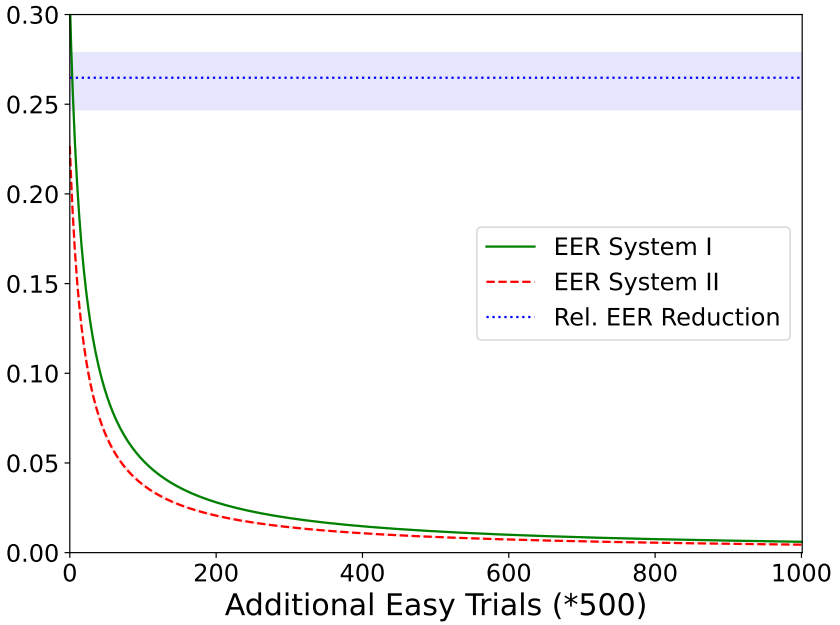}}
    \caption{EERs of System I and System II and relative EER reduction from System I to System II, with easy trials gradually added to the test. (a) Only negative easy trials are added; (b) Only positive easy trials are added;
    (c) Both negative and positive trials are added.}
  \label{fig:eer}
\end{figure}

By the simulated scores, we can compute the EER of each system and
the relative EER reduction from System I to System II.
To avoid any biased conclusion caused by randomness, we repeat this simulation 10 times and present
the mean and variance of the result EERs.
The results are shown in Fig.~\ref{fig:eer},
where we show the results when adding easy negative trials, easy positive trials, and both
in the three plots respectively.
It can be seen that the additional easy trials generate an obvious impact on system evaluation:
they not only reduce the EER value of each system (as expected)
but also impact the relative EER reduction.
Specifically, when only negative easy trials or positive easy trials are added
(i.e., Fig.~\ref{fig:eer}(a) and Fig.~\ref{fig:eer}(b) respectively),
the mean of the relative EER reduction is monotonically decreased.
This indicates that involving easy trials (either negative or positive) may results in
instability when performing system comparison (see the high variance),
and if the easy trials are in a large
amount, the relative EER reduction tends to be under-evaluated (see the declined mean).
This means that the discriminative power of the test could be diminished if many easy
trials are involved.

In particular, in the extreme case where
easy trials are infinite, any two systems tend to show the same
EERs, and the discriminative power of the test is totally lost.
This is not surprising, as any system can achieve an EER close to zero by setting a low (infinite
negative trials) or high (infinite positive trials) threshold.
In this extreme case, any effort in designing advanced techniques will be nullified.

We emphasize that the simulation is based on rather simple distributions.
For real data, the situation could be highly complex and other unexpected problems could be encountered.

\subsection{Identify hard trials}
\label{sec:hard-trials}

The analysis in the previous section demonstrated that involving a large number of easy trials may cause potential problems in system evaluation and comparison.
Unfortunately, most benchmarks are based on cross-pairing which produces lots of easy trials.
To solve the problem, we wish to eliminate easy trials from the test set and only retain hard trials.

In this section, we address the question `how to define and identify hard trials'.
A natural idea is to select the trials that the target system
tends to give incorrect decisions, however we cannot use the target
system to perform the selection, otherwise the error rate will be 100\%.
A possible way is to choose a committee of reference systems
and combine their decisions to select hard trials.

Motivated by the idea of margin loss~\cite{vapnik1963pattern,vapnik1982estimation,cortes1995support},
we propose an SVM-based approach to combine decisions of reference systems.
Specifically, for each trial, we collect the scores produced by the reference systems and form a score vector for that trial.
Then we train a support vector machine (SVM) with score vectors of all the trials in the test set, by setting the classification task as
discriminating the positive and negative trials. Once the SVM has been trained,
the trials corresponding to the \emph{support vectors} are selected as hard trials.

Defining hard trials in this way can find support from several theories. Firstly,
according to the theory of SVM, support vectors are samples on or outside of the margin. In other words,
they are close to the decision boundary and so are naturally hard for the classification of negative and positive trials, i.e., the speaker verification task.
Secondly, the SVM can be regarded as a representative model for existing speaker recognition techniques,
so the non-support vectors can be regarded as the test cases that existing techniques can well recognize.
In this sense, the SVM is a base model and new models can only focus on the cases that the SVM feels uncertain, i.e., the support vectors.
According to the boosting theory~\cite{ferreira2012boosting,schapire2013explaining},
we can always combine the SVM and the new models to gain better performance.

\section{Related work}

\noindent \textbf{Hard trials in existing benchmarks:} The idea of focusing on hard trials (at least not too easy ones) can be found in the design of several existing benchmarks.
For instance, the NIST SRE tried to boost difficulty of the tasks by
splitting gender-independent trials into gender-dependent trials
and/or grabbing multi-session data to produce multi/cross-session trials~\cite{martin2009nist}.
The VoxSRC 2019 evaluation also splits the
trials into easy and hard sets~\cite{nagrani2020voxceleb} according to speaker's nationality and gender.
However, the crucial role of hard trials for performance
evaluation and technical comparison is not widely recognized,
and how to choose hard trials has not been systematically studied.

\noindent  \textbf{Hard trials and margin loss:} The idea of hard trials is partly motivated by margin loss~\cite{vapnik1963pattern,vapnik1982estimation,cortes1995support},
by which only samples on or outside of the margin (support vectors) determine the model and incur loss. The margin
loss has been widely used in speaker recognition, e.g.,  metric learning with triple-loss~\cite{novoselov2018triplet}.
It was proved that margin loss offers good generalizability (precisely, a bounded error rate on test data)~\cite{vapnik1982estimation,cortes1995support}.
Margin loss is the core ingredient of SVM, and the support vectors are those that are uneasy to classify by the learned SVM therefore
can be regarded as hard. In this study, we borrow this idea to mine hard trials.

\noindent  \textbf{Trial selection and FA/FR trade-off:} Trial selection should not be confused with FA/FR trade-off. For FA/FR trade-off, the trials are fixed, though the
threshold changes to achieve different FA/FR balance. This trade-off determines the decision behavior of the verification system.
In contrast, trial selection aims to test the target system on different deployment conditions.
If the FA/FR trade-off is called `operation point selection', the trial selection is `operation environment selection'.

Nonetheless, trials are indeed related to FA/FR trade-off. For instance, the \emph{area under the ROC curve (AUC)} that reflects
the overall performance with various FA/FR trade-off is proportional to the number of `odd trial pairs', where each odd trial pair
involves a low-score positive trial and a high-score negative trial~\cite{garcia2012optimization}.
Trial selection was also used in partial AUC optimization~\cite{bai2019partial}.
In that work, the authors tried to optimize part of the AUC in order to gain better overall performance.
To achieve the goal, they selected a subset of negative trials, and combined them with the original positive trials to
compute the partial AUC loss. Their work, although applied to trial selection, is not related to perform evaluation.

\section{Experiment}
\label{sec:exp}

\subsection{Data}
\label{sec:data}

Two datasets were used in our experiments, VoxCeleb~\cite{nagrani2017voxceleb,chung2018voxceleb2} and SITW~\cite{mclaren2016speakers}.
More information is presented below.

\begin{itemize}

\item VoxCeleb\footnote{http://www.robots.ox.ac.uk/$\sim$vgg/data/voxceleb/}:
A large-scale audio-visual speaker dataset collected by the University of Oxford, UK.
In our experiments, the development set of VoxCeleb2 was used to train the x-vector models,
which contains 5,994 speakers in total and entirely disjoints from the VoxCeleb1 and SITW datasets.
Trials of Cleaned \emph{VoxCeleb1}, \emph{VoxCeleb1-E} and \emph{VoxCeleb1-H} were used to test the performance.
Note that the pairs of \emph{VoxCeleb1-H} are drawn from identities with the same gender and nationality, hence
harder to verify than those in \emph{VoxCeleb1-E}.

\item SITW: A standard evaluation dataset excerpted from VoxCeleb1, which consists of 299 speakers.
Trials of \emph{Dev.Core} and \emph{Eval.Core} were used for evaluation.

\end{itemize}

\subsection{Basic systems}
\label{sec:base}

In this study, we firstly follow the VoxCeleb recipe of the Kaldi toolkit~\cite{povey2011kaldi}
to build our i-vector~\cite{dehak2011front} and x-vector~\cite{snyder2018xvector} baselines\footnote{https://github.com/kaldi-asr/kaldi/tree/master/egs/voxceleb/}.
This basic recipe may not achieve the best performance on a particular dataset,
but has been demonstrated to be highly competitive and generalizable by many researchers with their own data and model settings.
We therefore consider that this recipe can represent a stable state-of-the-art technique.
Moreover, using this recipe allows others to reproduce our results easily. No data augmentation was used.

\begin{itemize}

\item i-vector: The i-vector model was built following the Kaldi VoxCeleb/v1 recipe.
The acoustic features comprise 23-dimensional MFCCs plus the log energy, augmented by first- and second-order derivatives,
resulting in a 72-dimensional feature vector.
Moreover, cepstral mean normalization (CMN) is employed to normalize the channel effect,
and an energy-based voice active detection (VAD) is used to remove silence segments.
The UBM consists of 2,048 Gaussian components, and
the dimensionality of the i-vector space is 400. For the back-end model,
LDA is firstly used to reduce dimensionality to 200, and then
PLDA~\cite{Ioffe06} is employed to score the trials.

\item x-vector: The x-vector model was created following the Kaldi VoxCeleb/v2 recipe. The acoustic features are 30-dimensional MFCCs.
The DNN architecture involves 5 time-delay (TD) layers to learn frame-level deep speaker features,
and a temporal statistic pooling (TSP) layer is used to accumulate the frame-level features to utterance-level statistics, including the mean and standard deviation.
After the pooling layer, 2 fully-connection (FC) layers are used as the classifier, for which the outputs correspond to
the number of speakers in the training set.
Once trained, the 512-dimensional activations of the penultimate layer are read out as an x-vector.
The back-end model is the same as in the i-vector system.

\end{itemize}

The EER and minDCF results with 2 baseline systems on 5 test sets are shown in Table~\ref{tab:base}.
The results demonstrate that modern speaker verification systems can obtain reasonable performance, even with the basic settings.

\begin{table}[htb!]
\centering
\caption{EER(\%) and minDCF results with 2 baseline systems (i-vector and x-vector) on 5 test sets.}
\label{tab:base}
\scalebox{0.86}{
\begin{tabular}{ccccccc}
\toprule
\multicolumn{7}{c}{EER\%} \\
\toprule
\textbf{System} & \textbf{Model} & \textbf{Vox1-O} & \textbf{Vox1-E} & \textbf{Vox1-H} & \textbf{SITW-D.C} & \textbf{SITW-E.C}  \\
\cmidrule(r){1-7}
1 & i-vector       &   5.819         &   5.872         &   9.536         &   5.545           &   6.315   \\
\cmidrule(r){1-7}
2 & x-vector       &   4.558         &   4.290         &   7.121         &   4.120           &   4.975   \\
\toprule
\multicolumn{7}{c}{minDCF (p-target = 0.01)} \\
\toprule
\textbf{System} & \textbf{Model} & \textbf{Vox1-O} & \textbf{Vox1-E} & \textbf{Vox1-H} & \textbf{SITW-D.C} & \textbf{SITW-E.C}  \\
\cmidrule(r){1-7}
1 & i-vector       &   0.5189        &  0.5038         &   0.6366        &   0.4504          &   0.5178   \\
\cmidrule(r){1-7}
2 & x-vector       &   0.4882        &  0.4343         &   0.5703        &   0.4292          &   0.4660   \\
\bottomrule
\end{tabular}}
\end{table}

\subsection{More powerful systems}
\label{sec:sota}

We also constructed more powerful x-vector systems to reproduce the performance of the SOTA speaker recognition techniques.
A bunch of state-of-the-art architectures/techniques, including ResNet34SE topology~\cite{he2016deep},
attentive statistics pooling~\cite{okabe2018attentive})
and angular margin-based training objectives (AM-Softmax~\cite{wang2018additive} and AAM-Softmax~\cite{deng2019arcface})
are employed in our experiments. Specifically, the acoustic features are 80-dimensional Fbanks.
The neural backbone adopts the ResNet34 topology for frame-level feature extraction.
The TSP and ASP strategies are used to construct utterance-level representations.
These representations are then transformed by a fully-connected layer to generate logits and are fed to a softmax
layer to generate posterior probabilities over speakers.
The training objective employs AM-Softmax and AAM-Softmax, and the margin factor is set to 0.2 and the scale factor is set to 30.
Once trained, the 256-dimensional activations of the last fully-connection layer are read out as an x-vector.
With the margin-based objectives, the simple cosine metric is used to score the trials.
No data augmentation is used.
The source code has been published online to help readers reproduce our systems\footnote{https://gitlab.com/csltstu/sunine}.

The EER and minDCF results of these more powerful systems are presented in Table~\ref{tab:sota}.
It can be observed that the SOTA systems can obtain rather good performance on all the test sets.
Although there is no complicated back-end model such as PLDA and auxiliary score normalization/calibration strategies,
the best EER results can still be less than 2.0\% in some test cases.

\subsection{Hard trials retrieval}
\label{sec:hard-mine}

We firstly built 8 baseline systems, including 4 i-vectors and 4 x-vectors,
to perform statistical analysis on PLDA scores.
For i-vectors, the number of Gaussian components of the universal background model (UBM) is set from 2048 to 256,
and the dimensionality of the i-vector is set to be 400 and 200.
For x-vectors, the unit number of each hidden layer is set from 1024 to 128,
and the dimensionality of the x-vector is set to be 512 and 256.
The results in terms of the equal error rate (EER) are reported in Table~\ref{tab:base2}.

\begin{table}[htb!]
\renewcommand{\thetable}{III}
\caption{EER(\%) results with 8 baseline systems on 5 test sets.
For i-vectors, X-Y represents the component number and i-vector dimension.
For x-vectors, X-Y represents the hidden unit number and x-vector dimension.}
\label{tab:base2}
\centering
\scalebox{0.86}{
\begin{tabular}{llccccc}
\toprule
\textbf{Model} & \textbf{Config} & \textbf{Vox1-O} & \textbf{Vox1-E} & \textbf{Vox1-H} & \textbf{SITW-D.C} & \textbf{SITW-E.C}  \\
\cmidrule(r){1-7}
i-vector    & 2048-400           &   5.834         &   5.899         &   9.437         &   5.853           &   6.725   \\
            & 1024-400           &   6.446         &   6.630         &   10.52         &   6.276           &   7.326   \\
            & 512-200            &   7.095         &   7.315         &   11.40         &   7.047           &   7.736   \\
            & 256-200            &   7.834         &   7.958         &   12.33         &   7.547           &   8.748   \\

\cmidrule(r){1-7}
x-vector    & 1024-512           &   5.116         &   4.858         &   7.633         &   6.546           &   7.682   \\
            & 512-512            &   5.356         &   5.086         &   7.970         &   6.662           &   7.354   \\
            & 256-256            &   5.111         &   4.930         &   8.051         &   6.585           &   7.709   \\
            & 128-256            &   5.967         &   5.797         &   9.239         &   7.008           &   8.475   \\
\bottomrule
\end{tabular}}
\end{table}

Hard trials are identified as follows:

\begin{itemize}
\item For each test trial, the PLDA scores from 8 baseline systems are concatenated into a 8-dimensional score vector.
\item The score vectors of all the trials in the test set are used to train an SVM model.
The simple linear SVM was employed in our experiment\footnote{https://scikit-learn.org/stable/modules/generated/sklearn.svm.SVC.html}.
\item The trials corresponding to the support vectors are identified as hard trials.
\end{itemize}

The numbers of identified hard trials in different test sets are shown in Table~\ref{tab:hard}.
It can be seen that the proportion of hard trials varies from 1/6 to 1/10,
reflecting the different levels of the challenge of each test set.
For instance, the hard trials in Vox1-H are nearly two times of those in Vox1-E,
demonstrating the recognition task on Vox1-H is more challenging.

\begin{table*}[htp!]
\renewcommand{\thetable}{II}
\caption{EER(\%) and minDCF results with more powerful systems on 5 test sets.}
\centering
\label{tab:sota}
\scalebox{1.0}{
\begin{tabular}{ccccccccccc}
\toprule
\multicolumn{11}{c}{EER\%} \\
\toprule
\textbf{System} & \textbf{Backbone} & \textbf{Pooling} & \textbf{Loss} & \textbf{Dim} & \textbf{Backend} & \textbf{Vox1-O} & \textbf{Vox1-H} & \textbf{Vox1-E} & \textbf{SITW-D.C} & \textbf{SITW-E.C}  \\
\cmidrule(r){1-6} \cmidrule(r){7-9} \cmidrule(r){10-11}
3  & TDNN       & TSP  & AM-Softmax   & 512  & Cosine  & 3.478  & 5.821  & 3.564  & 6.276  & 6.534  \\
4  & RestNet34  & TSP  & AM-Softmax   & 256  & Cosine  & 1.633  & 2.857  & 1.688  & 3.119  & 3.253  \\
5  & RestNet34  & TSP  & AAM-Softmax  & 256  & Cosine  & 1.803  & 2.892  & 1.747  & 2.926  & 3.253  \\
6  & RestNet34  & ASP  & AM-Softmax   & 256  & Cosine  & 1.489  & 2.398  & 1.412  & 2.233  & 2.488  \\
\toprule
\multicolumn{11}{c}{minDCF (p-target = 0.01)} \\
\toprule
\textbf{System} & \textbf{Backbone} & \textbf{Pooling} & \textbf{Loss} & \textbf{Dim} & \textbf{Backend} & \textbf{Vox1-O} & \textbf{Vox1-H} & \textbf{Vox1-E} & \textbf{SITW-D.C} & \textbf{SITW-E.C}  \\
\cmidrule(r){1-6} \cmidrule(r){7-9} \cmidrule(r){10-11}
3  & TDNN       & TSP  & AM-Softmax   & 512  & Cosine  & 0.3558 & 0.4968 & 0.3779 & 0.4948 & 0.5905  \\
4  & RestNet34  & TSP  & AM-Softmax   & 256  & Cosine  & 0.1771 & 0.2673 & 0.1899 & 0.2453 & 0.2812  \\
5  & RestNet34  & TSP  & AAM-Softmax  & 256  & Cosine  & 0.1961 & 0.2796 & 0.1946 & 0.2474 & 0.2729  \\
6  & RestNet34  & ASP  & AM-Softmax   & 256  & Cosine  & 0.1341 & 0.2249 & 0.1540 & 0.1818 & 0.2109  \\
\bottomrule
\end{tabular}}
\end{table*}

\begin{table*}[htb!]
\renewcommand{\thetable}{V}
\caption{EER(\%) and minDCF results with SOTA systems on the hard trials identified by the SVM-based approach. \newline (H) denotes the \textbf{H}ard trials of the corresponding test set.}
\centering
\label{tab:hard-svm}
\scalebox{1.0}{
\begin{tabular}{ccccccccccc}
\toprule
\multicolumn{11}{c}{EER\%}\\
\toprule
\textbf{System} & \textbf{Backbone} & \textbf{Pooling} & \textbf{Loss} & \textbf{Dim} & \textbf{Backend} & \textbf{Vox1-O(H)} & \textbf{Vox1-H(H)} & \textbf{Vox1-E(H)} & \textbf{SITW-D.C(H)} & \textbf{SITW-E.C(H)}  \\
\cmidrule(r){1-6} \cmidrule(r){7-9} \cmidrule(r){10-11}
3  & TDNN       & TSP  & AM-Softmax   & 512  & Cosine   & 23.453 & 26.239 & 24.529 & 33.046 & 33.630 \\
4  & RestNet34  & TSP  & AM-Softmax   & 256  & Cosine   & 11.641 & 13.517 & 12.625 & 17.241 & 17.082 \\
5  & RestNet34  & TSP  & AAM-Softmax  & 256  & Cosine   & 12.550 & 13.671 & 13.153 & 16.092 & 17.082 \\
6  & RestNet34  & ASP  & AM-Softmax   & 256  & Cosine   & 11.073 & 11.696 & 10.978 & 13.506 & 13.701 \\
\toprule
\multicolumn{11}{c}{minDCF (p-target = 0.01)}\\
\toprule
\textbf{System} & \textbf{Backbone} & \textbf{Pooling} & \textbf{Loss} & \textbf{Dim} & \textbf{Backend} & \textbf{Vox1-O(H)} & \textbf{Vox1-H(H)} & \textbf{Vox1-E(H)} & \textbf{SITW-D.C(H)} & \textbf{SITW-E.C(H)}  \\
\cmidrule(r){1-6} \cmidrule(r){7-9} \cmidrule(r){10-11}
3  & TDNN       & TSP  & AM-Softmax   & 512  & Cosine   & 0.9779 & 1.0000 & 0.9999 & 1.0000 & 1.0000 \\
4  & RestNet34  & TSP  & AM-Softmax   & 256  & Cosine   & 0.7964 & 0.8998 & 0.8894 & 0.9242 & 0.9683 \\
5  & RestNet34  & TSP  & AAM-Softmax  & 256  & Cosine   & 0.8455 & 0.9028 & 0.8837 & 0.9522 & 0.9526 \\
6  & RestNet34  & ASP  & AM-Softmax   & 256  & Cosine   & 0.7210 & 0.8191 & 0.8143 & 0.8134 & 0.8565 \\
\bottomrule
\end{tabular}}
\end{table*}

\begin{table}[htb!]
\renewcommand{\thetable}{IV}
\caption{Hard trials retrieved by the SVM-based approach.}
\label{tab:hard}
\centering
\scalebox{0.92}{
\begin{tabular}{lcccc}
\toprule
\multirow{2}{*}{Trials} & \multicolumn{2}{c}{Origin} &   \multicolumn{2}{c}{Hard Trials}  \\
             & \# of Target  & \# of Non-target  &  \# of Target  &  \# of Non-target  \\
\cmidrule(r){1-1} \cmidrule(r){2-3}  \cmidrule(r){4-5}
Vox1-O          &  18,802    &  18,809      &   1,761   &  1,766  \\
Vox1-E          &  289,921   &  289,897     &   28,183  &  28,182  \\
Vox1-H          &  275,488   &  275,406     &   43,492  &  43,499  \\
\cmidrule(r){1-1} \cmidrule(r){2-3}  \cmidrule(r){4-5}
SITW-D.C        &  2,597     &  335,629     &  348     &  44,659   \\
SITW-E.C        &  3,658     &  718,130     &  562     &  109,671  \\
\bottomrule
\end{tabular}}
\end{table}

\subsection{Performance on hard trials}
\label{sec:hard-res}

The EER and minDCF results on hard trials are shown in Table~\ref{tab:hard-svm}.
Note that results on the full origin trials have been presented in Table~\ref{tab:sota}.
It can be observed that compared to the results on the full trials,
the results on the set of hard trials are drastically reduced in terms of both EER and minDCF,
indicating that the hard trials selected from the baseline systems are truly hard,
even for systems employing advanced techniques.
It means that some trials are commonly-agreed hard and speaker recognition performance is still far
from satisfactory on these trials, even with the SOTA techniques.

Another observation is that the results on hard trials amplify discrepancy among systems and techniques, in terms of absolute EER/minDCF values.
The relative EER/minDCF among systems/techniques does not show much difference when tested on
full trials and hard trials, though the values obtained on full trials are slightly higher.
This is good news as it shows that using the full trials did not cause serious problems so far. However, the test is only for techniques that have been widely confirmed (e.g., ResNet, ASP);
how other techniques behave on
full trials and hard trials is unknown and will be our future work.

\section{Conclusions}

In this paper, we advocated paying attention to hard trials.
The issue was originally raised by the overwhelming easy trials
generated by the cross-pairing scheme which has been widely-adopted by speaker recognition benchmarks when designing the test protocol.
We demonstrated by theoretical analysis and
simulation experiments that vast easy trials may lead to biased results in performance evaluation and system comparison.
To solve the problem, we proposed to focus on hard trials and
presented an SVM-based identification approach.
We employed the proposed approach to identify hard trials in two famous test sets (VoxCeleb1 and SITW)
and conducted a thorough performance evaluation.
The results demonstrated that some trials are truly hard and
the present performance obtained by SOTA technologies on the hard trials is far from satisfactory.

Although our analysis showed that easy trials may cause under-evaluated relative EER,
this seems not the case in our tests performed on VoxCeleb and SITW,
at least with the selected techniques in the experiment. More investigate
is certainly required to discover the reason, but we highly recommend researchers
testing their new methods on hard trials, in order to avoid any under evaluation.

\section*{Acknowledgment}

This work was supported by the National Natural Science Foundation of China (NSFC) under Grant No.62171250.


\bibliographystyle{IEEEtran}
\bibliography{mybib}

\end{document}